%% file: main.tex
\documentclass[11pt,twocolumn]{scrartcl}

\usepackage{casa_conf}	
\usepackage{graphicx}	

\usepackage{amsmath}
\usepackage{tabularx, makecell}
\usepackage{xcolor}
\usepackage{ulem}
\usepackage{wrapfig}
\usepackage[format=plain]{caption}
\usepackage{hyperref}

\newcommand{\modify}[2]{#2}

\newcommand\etal{\textit{et al.~}}

\newenvironment{modified}
    {
    }
    {}

\title{Learning-based pose edition for efficient and interactive design}

\author{Léon Victor$^{1,3}$, 
        Alexandre Meyer$^{1,2}$, 
        Saïda Bouakaz$^{1,2}$
        
        \\
        leon.victor@insa-lyon.fr, \{alexandre.meyer, saida.bouakaz\}@univ-lyon1.fr
       }
       
\date{%
    $^1$ Univ Lyon, LIRIS, UMR CNRS 5205\\%
    $^2$ Université Claude Bernard Lyon 1\\%
    $^3$ INSA Lyon\\%
    \vspace{5pt}
}

\usepackage{capt-of, etoolbox}
\begin{document}
\makeatletter
\let\@oldmaketitle\@maketitle
\renewcommand{\@maketitle}{\@oldmaketitle
  \vspace{-10mm}
  \centering
  \includegraphics[width=0.8\linewidth]{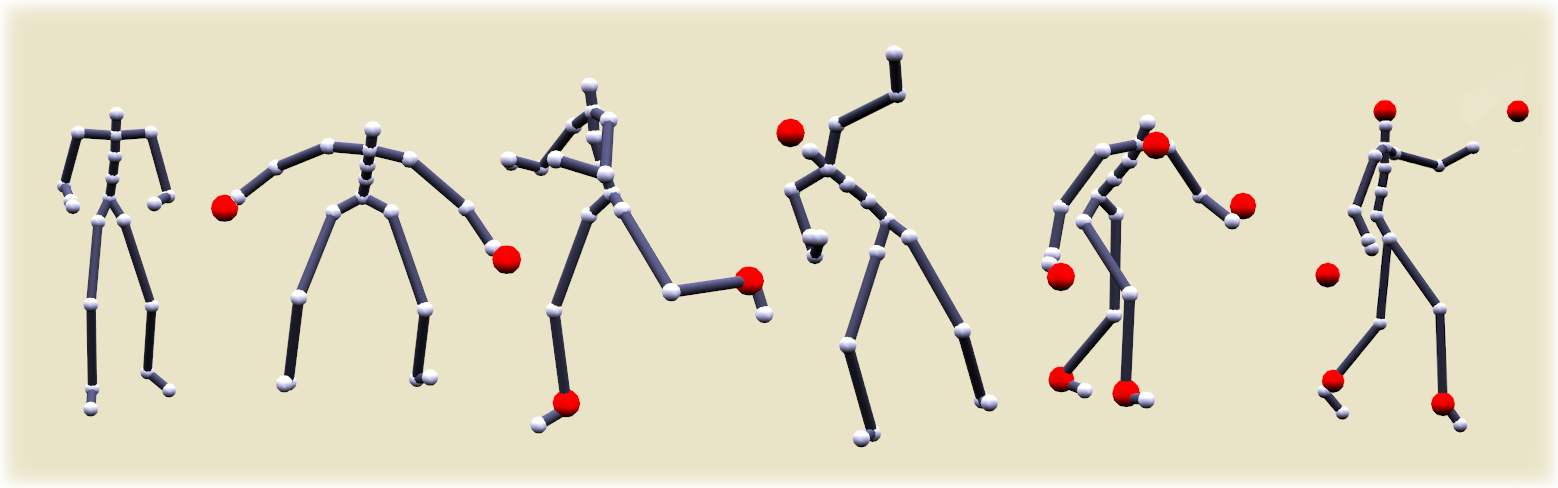}
  \captionof{figure}{Sample results of our method in various configuration. Our method is able to generate plausible poses given a starting pose (on the left) and some targets (in red), respecting skeleton constraints without having to explicitly specify them.}
    \bigskip}
\makeatother
\vspace{-5pt}
\maketitle
\vspace{4pt}
\begin{abstract}
Authoring an appealing animation for a virtual character is a challenging task. In computer-aided keyframe animation artists define the key poses of a character by manipulating its underlying skeletons.
To look plausible, a character pose must respect many ill-defined constraints, and so the resulting realism greatly depends on the animator's skill and knowledge.
Animation software provide tools to help in this matter, relying on various algorithms to automatically enforce some of these constraints.
The increasing availability of motion capture data has raised interest in data-driven approaches to pose design, with the potential of shifting more of the task of assessing realism from the artist to the computer, and to provide easier access to non-experts. In this paper, we propose such a method, relying on neural networks to automatically learn the constraints from the data. We describe an efficient tool for pose design, allowing naïve users to intuitively manipulate a pose to create character animations.
\footnote{The code is available at \url{https://github.com/leonvictor/neural-pose-edition}}
\linebreak
\linebreak
\end{abstract}

\section{Introduction}
Character animation is an essential part of computer-generated imagery industries such as feature films, cartoons or video games which make use of on-screen characters to tell stories, convey emotions and appeal to their audiences.
These characters are represented by 3-dimensional meshes whose motion is driven by an underlying skeleton. A common method to design and edit animations is keyframing: animator pose the character at desired time stamps (the key frames) and the computer interpolates between them to fill in the gaps.
Most animation software such as Blender or Maya provide interactive tools allowing users to pose a character by manipulating its underlying skeleton.
We propose an innovative solution that makes the pose editing process more affordable without compromising the quality of the results. The presented method leverages neural networks to implicitly learn the intricacies of a (human) skeleton and provide simple controls.
Our main goal is to create an intuitive real-time system that can produce appealing poses even for a novice user.

Our framework relies on a few small networks requiring reasonable resources to train, with the added advantage of running quite fast at inference time. The core of our approach is an encoder-decoder trained on skeleton pose data, the task of which is to build a latent representation of the pose space, alleviating some of the limitations of the former. We then train a family of solver networks to work on this latent space in order to generate a pose satisfying user-defined target positions.
\section{Related work}
\label{sect:related_work}
\input{related_work}
    
\section{Proposed method}
\label{sect:proposedMethod}
\input{proposed_method}
\section{Results}
\label{sect:results}
\input{results}


\section{Conclusion and perspectives}
\label{sect:conclusion}
\input{conclusion}

\bibliographystyle{unsrt}
\bibliography{bibliography}

\end{document}

%% file: related_work.tex
The industry standard for pose edition is to create rigs, a collection of pieces of software designed to manipulate a character's skeleton. The rig describes the skeleton's bones, how they relate to each other, are constrained in their possible motion and are deformed. These rules are loosely specified and creating a good rig requires a detailed understanding of physics and anatomy, as well as technical and artistic skills. Rigging is thus a time consuming task even for experienced animators, and even more so in large scale productions which often require a different in-depth rig for each character in the cast.
Previous work has helped alleviate this difficulty by providing efficient tools to speed up/and or ease the rigging process, relying on inverse kinematics or data-driven methods.
\subsection{Character pose design}
\subsubsection{Inverse Kinematics (IK)}
IK solvers are a family of methods commonly used in robotics, engineering and computer graphics, in which the parameterization of a kinematic chain is determined from the position of its end effector.
They are a staple tool in pose design software, ensuring the respect of elementary constraints during pose edition. Their de-facto role is to guarantee the length of the limbs, and in some cases to enforce the orientation angle range of a joint.
Many IK solutions have been studied over the years \cite{aristidou_inverse_2018}; usually revolving around approximated linearizations or heuristics. 

Numerical methods require a set of iterations to achieve a satisfactory solution formulated by a cost function to be minimized.
IK solutions can generally be divided into three sub-categories: Jacobian \cite{Siciliano_Handbook_Robot_2007}, Newtonians \cite{cohen_ik_1996} and Heuristics. Most software implement heuristic methods such as Cyclic Coordinate Descent (CCD) \cite{wang_ccd_1991} or 
Forward-Backward Reaching IK (FABRIK) \cite{aristidou_fabrik:_2011} due to their simplicity and extensibility. 

The main drawback of 
these solvers is that they manipulate kinematic chains without taking into account many morphological aspects that make a pose more or less plausible. They offer a first level of help to users but are not sufficient to guarantee a realistic pose. Many joints constraints are dependent on each other and require subjective, human-made approximations.

\subsubsection{Data-driven pose edition}
Data-driven methods offer promising opportunities to solve these approximations. Using real-life data can help in modelling the complex inter-dependencies of skeletons and providing users with smarter edition tools.
While it is still an early field of research, some solutions have been studied. Wu \etal \cite{wu_posing_2009} propose a method for natural character posing from a large motion database. It employs adaptive KD-clustering to select a representative frame from a database and sparse approximations to accelerate training and posing. 
Huang \etal in \cite{Huang_IK_MGDM_2017} present a method based on the formulation of multi-variate Gaussian distribution models (MGDMs), which learn the joint constraints of a kinematic skeleton from motion capture data. 

Some work has also been dedicated to finding new editing interfaces. \modify{}{Instead of the usual setup manipulating joints directly, Guay \etal \cite{guay_line_2013} articulate a framework based on the conceptual "line of action" which describes the overall pose dynamics. They provide a mathematical definition of the line of action, and a interface in which the software modifies the pose to follow a user-provided line. In the same line of though} Garcia \etal \cite{garcia_sketching_2019} propose \modify{a method transforming doodle of trajectories (position and orientation over time) }{a virtual reality-based interface where the user's hands motion (position and orientation over time) are transformed} into sequences of actions and then into detailed character animations using a dataset of parametrized motion clips automatically fitted to the trajectory. 

\subsection{Neural modelling of human motion}
Neural networks have received a great amount of attention over the last decade and shown impressive result in modelling complex data. Human motion has not been spared and deep learning methods have proven their capability of generating realistic motion in a number of difficult cases. 

The literature in neural-based animation include example in user-controlled character navigation \cite{Holden2017} and interactions with the environment \cite{starke_neural_2019}. 
Holden \etal \cite{Holden2020} also show that neural networks can be used to replace parts of existing data-driven methods, improving their scalability potential.
More recently, some work has also focused on improving smaller parts of the animation pipeline rather than replacing it completely. Berson et al. \cite{berson_intuitive_2020} leverage neural networks to provide an interactive system to edit facial animation. 

Data-driven IK and pose editing can relieve animators from time-consuming, back-and-forth pose adjustments by applying constraints extracted from real-world data. Recently, neural-network-based approaches have demonstrated their ability to model the intricacies of human motion while scaling to large amount of data and retaining a fast inference time. In this paper we seek to take advantage of these properties to create an efficient posing tool, intuitively usable even by a inexperienced user.

%% file: proposed_method.tex
\label{sect:proposed_method}
\subsection{Method overview}

\begin{figure*}[h]
    \centering
    \includegraphics[width=0.75\linewidth]{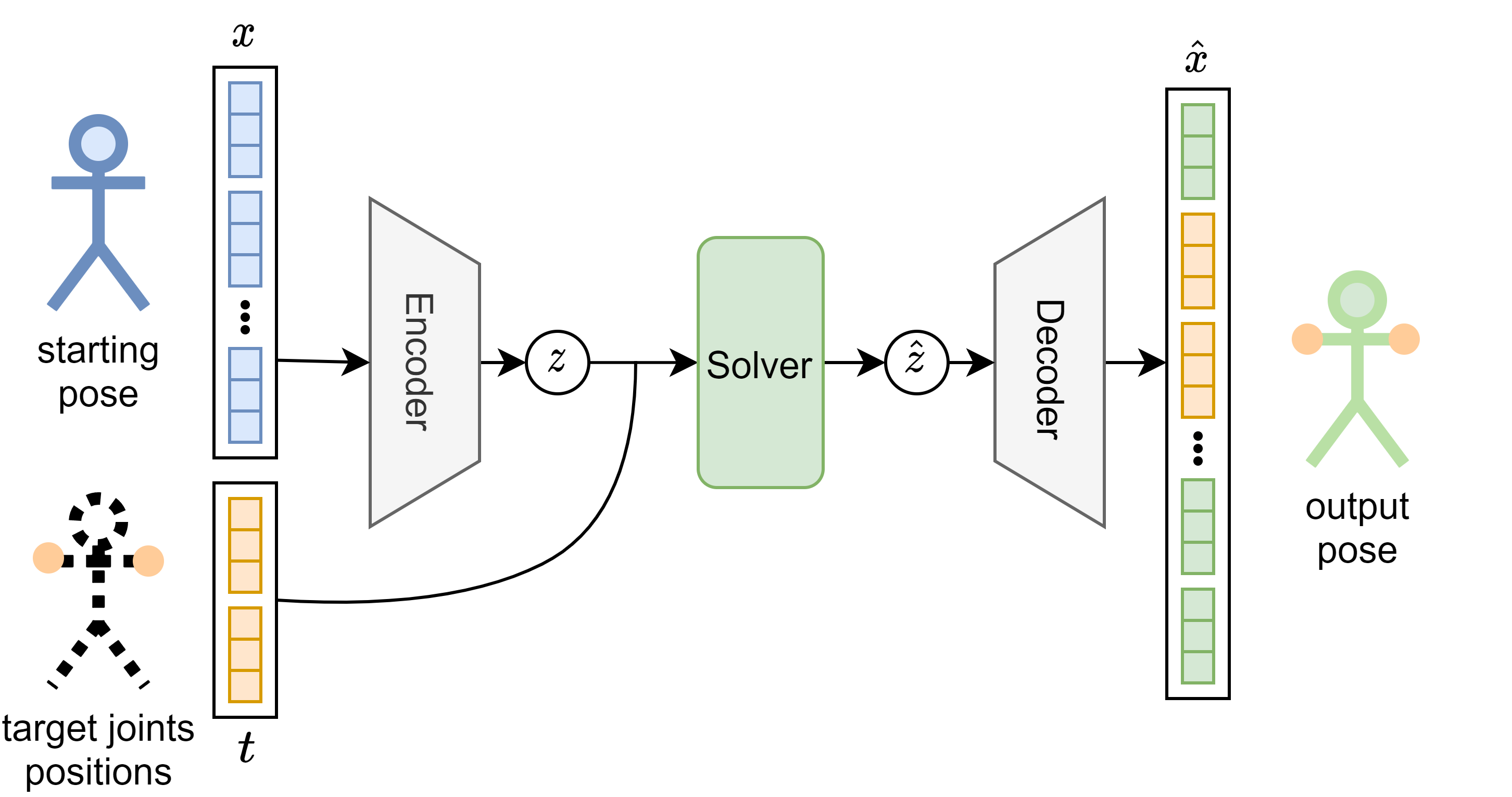}
    \caption{High level overview of the generation setup. The target joint's positions (yellow) are matched as closely as possible, while the other joints (green) should be as close as possible to the starting pose (blue).}
    \label{fig:generation_setup}
\end{figure*}

We propose a method to solve a high level pose design problem in which a pose is modified to reach desired target positions for some of its joints. We leverage the modelling power of neural networks to implicitly learn skeleton constraints from a pre-existing pose database.
Our method, illustrated in Fig. \ref{fig:generation_setup}, relies on two models: an auto-encoder to build an alternative latent pose space, and a solver model operating on this space to solve the pose design problem. 
We also describe an optional post-processing step to smooth out the remaining errors, and \modify{}{outline} a methodology using multiple instances of the solver model at once to work with a varying amount of targets.

\subsection{Data}
\label{sect:materiel}
    
We train the models using a dataset of human poses, obtained by processing multiple available motion-capture datasets from the literature: Emilya \cite{fourati_emilya_2014}, CMU \cite{CMU_BVH}, and the clips from Edinburgh university \cite{holden_deep_2016}. Each animation clip is retargeted to a standard skeleton following the scheme proposed by \cite{HoldenAE2015}. The global translation is removed, and each joint's position is calculated relative to the root joint, which is the projection of the pelvis on the floor. The unified skeleton is composed of 21 joints; using the joints' positions in space, a pose is described by $ 3 \times 21 = 63 $ float values concatenated in a single vector. The dataset is then formed by the individual poses in each clip. Before feeding them to the network we also normalize each pose by subtracting the mean and dividing by the standard deviation of each feature. \modify{}{With a few jittery clips manually removed, the final dataset used in the following experiments is composed of about 1,5 million poses.}

\subsection{Models description}
\subsubsection{Autoencoder}
Auto-encoders are made up of two neural networks tasked to learn efficient encodings of complex data. The encoder maps real data points to a learned, usually more compact, latent space; and the \modify{encoder}{decoder} maps them back to the original data space.
We build such an auto-encoder of poses in order to build a common operating space for the following solvers. Generating points in the latent space allows us to ensure that the output is always a plausible pose, as the decoder is trained to turn any and all latent point into them.

The encoder network is composed of two fully connected layers with \modify{195}{200} neurons and ReLU \cite{relu_2010} activations, followed by an output layer with no activation. The output layer's size is based on the number of
dimensions $d$ in which the latent representations are encoded. We empirically find that $d=64$ yields a good balance of representation accuracy and inference speed. The decoder is the exact reversed replica and uses the same set of weights. 

The autoencoder's weights are optimized by minimizing the mean squared error (MSE) between the input pose $x$ and its reconstructed equivalent $\hat x$ (Eq. \ref{eq:loss-ae}). In the following sections we refer to the encoder as $E$, the decoder as $D$ and a latent encoding as $z$, i.e. $z = E(x)$ and $\hat x = D(z)$. 

\begin{equation}
    \label{eq:loss-ae}
    L_{ae} = MSE(x, \hat x) = \frac{1}{d} \sum_{i=1}^{d}(x_i - \hat x_i)^2
\end{equation}

The autoencoder is trained for 20 epoch with batches of 256 poses, using the Adam optimizer \cite{kingma_adam_2017} with a learning rate of $0.0001$.

\subsubsection{Pose solver}

An instance of the solver model $S_t$ is specialized to solve the IK problem for $n$ specific targets $t$ and is trained to generate a new pose from an input pose and the desired targets locations. 
As it operates on the latent space built by the autoencoder, it more precisely accepts and outputs a latent pose vector, i.e. with $p_t$ the \modify{}{concatenated} target positions, $\hat z = S_t(z, p_t)$.

The network is composed of three fully connected layers with 126 neurons and ReLu activations, and an output layer with $d$ neurons.

During training, we randomly sample an input pose $x$ from the dataset and feed it to the network. \modify{The targets are generated by taking the positions of the considered joints on a random pose in the same animation clip as the input pose.}{We also sample a second pose $x'$ from the same source clip to use as target.} We found that this association helped the network learning by not relying on random (and possibly unreachable) target positions.

\modify{Its}{The network's} weights are optimized to minimize the loss function in Eq.\ref{eq:loss-ik} designed to represent its high level objective: \modify{reaching the targets with the associated joints while staying as close to the starting pose as possible}{reaching the targets with the associated joints while retaining a realistic pose}. We guide the network toward this objective by using a modified mean squared error function \modify{Eq.\ref{eq:loss-ik}}{, separating the poses ($x$ in this example) in two sets of joints: $x_{target}$ the joints associated to the targets $t$, and $x_{rest}$ the others} \modify{The generated pose's target joints $\hat x_{target}$ should be close to the input targets $p_t$, while its other joints $\hat x_{rest}$ should minimize their motion}{}.
We introduce a constant $k$ to give more relative importance to the target term of the function\modify{}{, so that the non-targets joints of $x'$ are only used to nudge the final result toward a plausible pose}. In our experiments $k$ is set to $0.01$.\modify{A side effect of our loss function is that the target positions are not an absolute truth to be reached at all cost. The solver is rather encouraged to use them as guides, only reaching them precisely when the starting pose would not require too much of a change.}{}

\begin{equation}
    \label{eq:loss-ik}
    L_{s} = MSE(\hat x_{target}, x'_{target}) + k \cdot MSE(\hat x _{rest}, x'_{rest})
\end{equation}

An instance of the solver model is trained for \modify{15}{5} epochs with the Adam optimizer using a learning rate of 0.0001 and a batch size of 256.

\subsection{Post processing}

It is a common observation with neural networks working with joints position that the generated positions can be jittery, and the resulting poses can suffer from slight variations in bone lengths. 
Our model is no exception, and while the variation is not visually detectable most of the time, computing the total bone length difference between the input skeleton and the generated pose shows that it is present. These variations are naturally undesirable and can result in visual discomfort on the spectator's end. In order to alleviate the problem we apply an optional post-processing step to the resulting poses to ensure constant bone lengths. We use the backward step from FABRIK as it is very lightweight computation-wise. Our experiments show that following this process  lends better results at a small cost in computing time (see table \ref{table:results}).

\subsection{Solving other targets configurations}
\label{sect:multi-solver}
Even though our solvers are designed to generate a pose considering one to two targets at once, it is possible to use multiple instances side by side and to switch to the correct one with regard to the selected targets. In cases where the user desires to use an arbitrary number of targets
(to suggest a position for a fixed joint for example) we can combine the multiple instances by running them in sequence, i.e. $\hat z = (S_{t3} \circ S_{t2} \circ S_{t1})(z)$ for $t1, t2, t3$ various targets and $S_{ti}$ the solvers trained to reach them.

%% file: results.tex

In order to evaluate the results of our method we integrate our solver in an example posing software and compare its outcome with a comparable, non-neural method: FABRIK \cite{aristidou_fabrik:_2011}. We pick FABRIK for the traits that make it a popular IK solver: its simplicity and fast convergence times. We implement a full-body human skeleton solver as described in \cite{aristidou_extending_2016} but stay as close as possible to our method setup process by not manually implementing any joint orient constraints.

\subsection{Visual results}
\begin{figure}[h]
    \centering
    \includegraphics[width=0.45\textwidth]{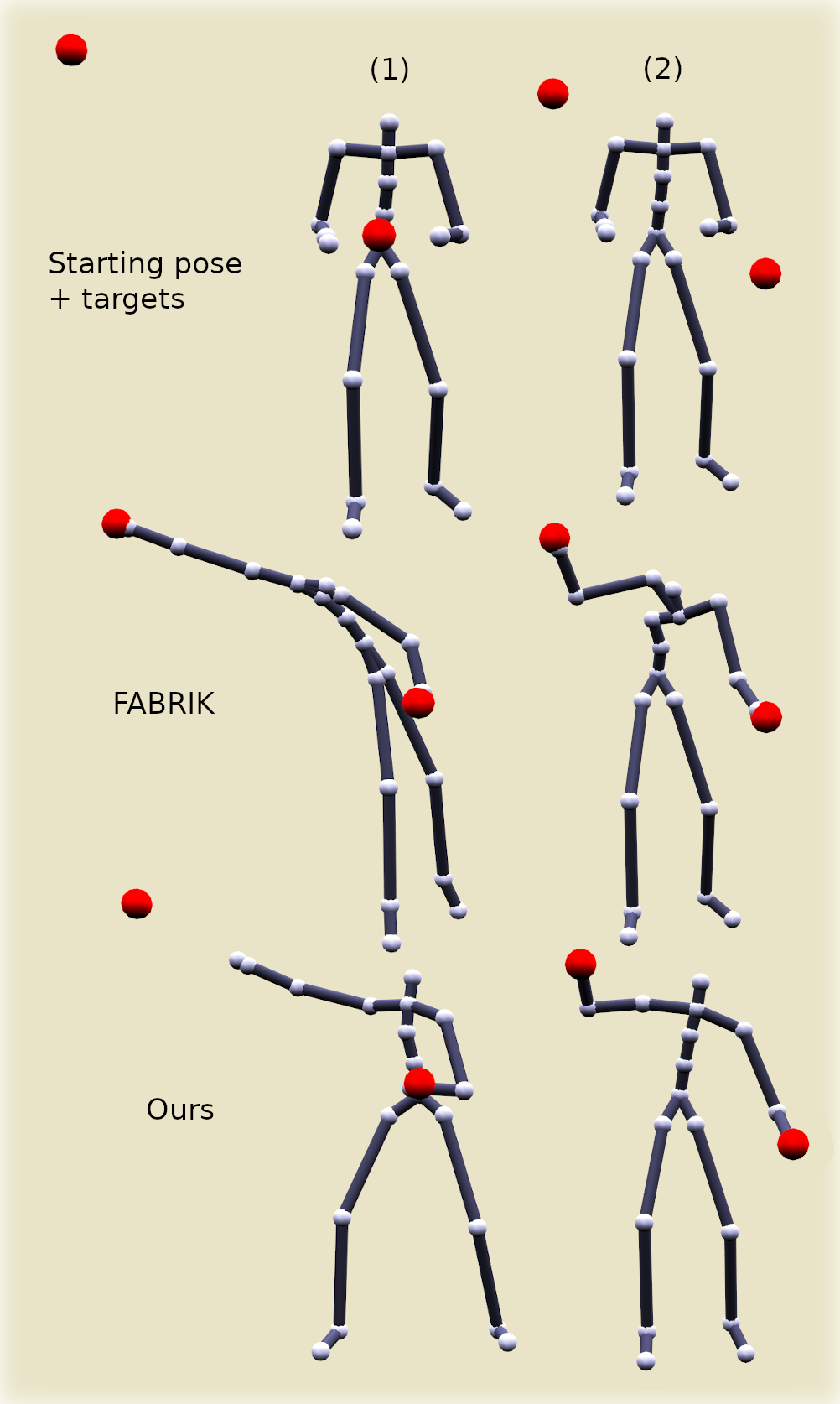}
    \caption{Starting from a pose and targets for two joints, an IK solver like FABRIK (middle) generates less realistic poses than our neural solver (bottom).}
    \label{fig:results-hands}
    \vspace{-15pt}
\end{figure}

Fig. \ref{fig:results-hands} showcases an example of how our method can be used to edit a pose by moving the targets around. in this case a single solver with the targets set to both skeleton's hands is used.

Our solver yields poses satisfying the constraints without breaking the implicit skeleton rules: the distance between limbs is constant, self-occlusion is avoided and the poses appear natural.
The side-by-side comparisons with FABRIK's results highlight the limits of working on kinematics chains with no prior on the human skeleton.

Example $(1)$ illustrates how the targets are used as guides rather than fixed, unbreakable rules. While FABRIK extends the full body, our solver generates a new pose where the torso is slightly twisted towards the right-hand target while the legs are spread to mimic maintaining balance.
Even though our method is aimed toward beginner animators, experienced ones could also find it useful. \modify{The results may lack precision but could a be sued}{It could for example be used} as a fast prototyping tool to flesh out the pose, while switching to more accurate and manipulation-heavy tools to focus on the details later on.

Examples of real-time usage of our method can also be found in the accompanying video.

\subsection{Combining solvers}

\begin{figure}[h!]
    \centering
    \includegraphics[width=0.9\linewidth]{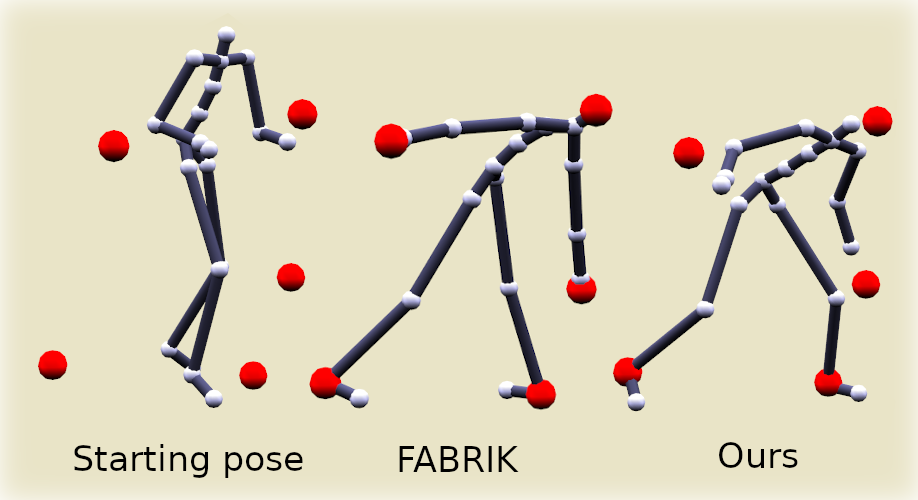}
    \caption{Sample results solving multiple targets with a sequence of neural solvers. Targets are shown in red. 
    }
    \label{fig:multi-solvers}
\end{figure}

Figure \ref{fig:multi-solvers} demonstrates an example with the multi-solver setup described in \ref{sect:multi-solver}. In this example three solvers are used at once: for the two hands, the two ankles and the head. Compared to the FABRIK result, our method yields a plausible pose: the skeleton is bent down to meet the head target, but the general orientation of the pose is kept intact. The limbs also retain some sort of curvature rather than fully extending in an unnatural way. Here again some of the targets are not strictly reached, as the pose generated by earlier solvers in the chain are modified by the others further down, but the guidance provided by the targets is respected. This setup also incurs slightly slower runtimes (see Table \ref{table:results}) but is still faster than FABRIK.
\subsection{Run times}

At run-time the complexity of the solver is fixed and regardless of the targets' positions, a single pass through the networks, which can be seen as just a few matrix multiplications, is enough to produce a result pose. This property coupled with the relatively small size of the networks allow for a fast solving process, as highlighted in table \ref{table:results}.

\begin{table}
    \centering
    \begin{tabularx}{\linewidth} {
         >{\raggedright\arraybackslash}X
        | >{\centering\arraybackslash}X
        | >{\centering\arraybackslash}X
    }
        \hline
        Method & Memory footprint (kB) & Runtime (ms)**\\
        \hline
        FABRIK (2)  & -   & 6.56     \\
        Ours (2)     & 442 & 1.47 (3.03*) \\
        FABRIK (5)  & -   & 6.74 \\
        Ours (5)     & 826 & 3.36 (4.58*) \\
        \hline
    \end{tabularx}
    \\
    \vspace{2mm}
    *With post-processing \\
    **Average over 1000 random iterations
    \caption{Comparative numeric results of the neural and FABRIK solvers with two and five end-effectors (using the combined solver method). All experiments are run on a single CPU thread.}
    \label{table:results}
    \vspace{-15pt}
\end{table}

Compared to other data-driven pose methods, the computing-heavy part of our process is done once at training time. Even so, the training itself is kept short thanks to the modest size of the networks: around an hour for the auto-encoder and 15 minutes for the solvers, on a single GPU.

\subsection{Memory footprint}
An advantage of neural networks is the low memory footprint they hold. While other data-driven pose design methods require the pose database (or a compressed version of it) to be kept in memory, neural networks only require their trained weights. These can be quite heavy as well in the case of large models, but as ours are quite small, so are their weights. As a comparison point, \cite{wu_posing_2009} discloses a 30MB memory footprint while our full-body solver only takes up 826kB.

\begin{modified}
\subsection{Comparison with other pose edition approaches}

Huang et al. \cite{Huang_IK_MGDM_2017} proposed a general comparison chart for full-body IK methods, ranking common approaches by speed and subjective quality. Adding our solution to the chart (Fig. \ref{fig:comparisons}) highlights the useful spot it fills by striking a good balance between speed and accuracy. 
To the best of our knowledge, this work presents the first method leveraging neural networks for pose edition. It stands apart from previous learning-based approaches as the first one to combine real-time edition speed with fully learned skeleton constraints. In comparison, NAT-IK \cite{Huang_IK_MGDM_2017} uses soft learned constraints but still requires explicit, manual ones to be set. \cite{wu_posing_2009} does not, but the poses are not generated in real time. 

\begin{figure}[h]
    \centering
    \includegraphics[width=0.5\textwidth]{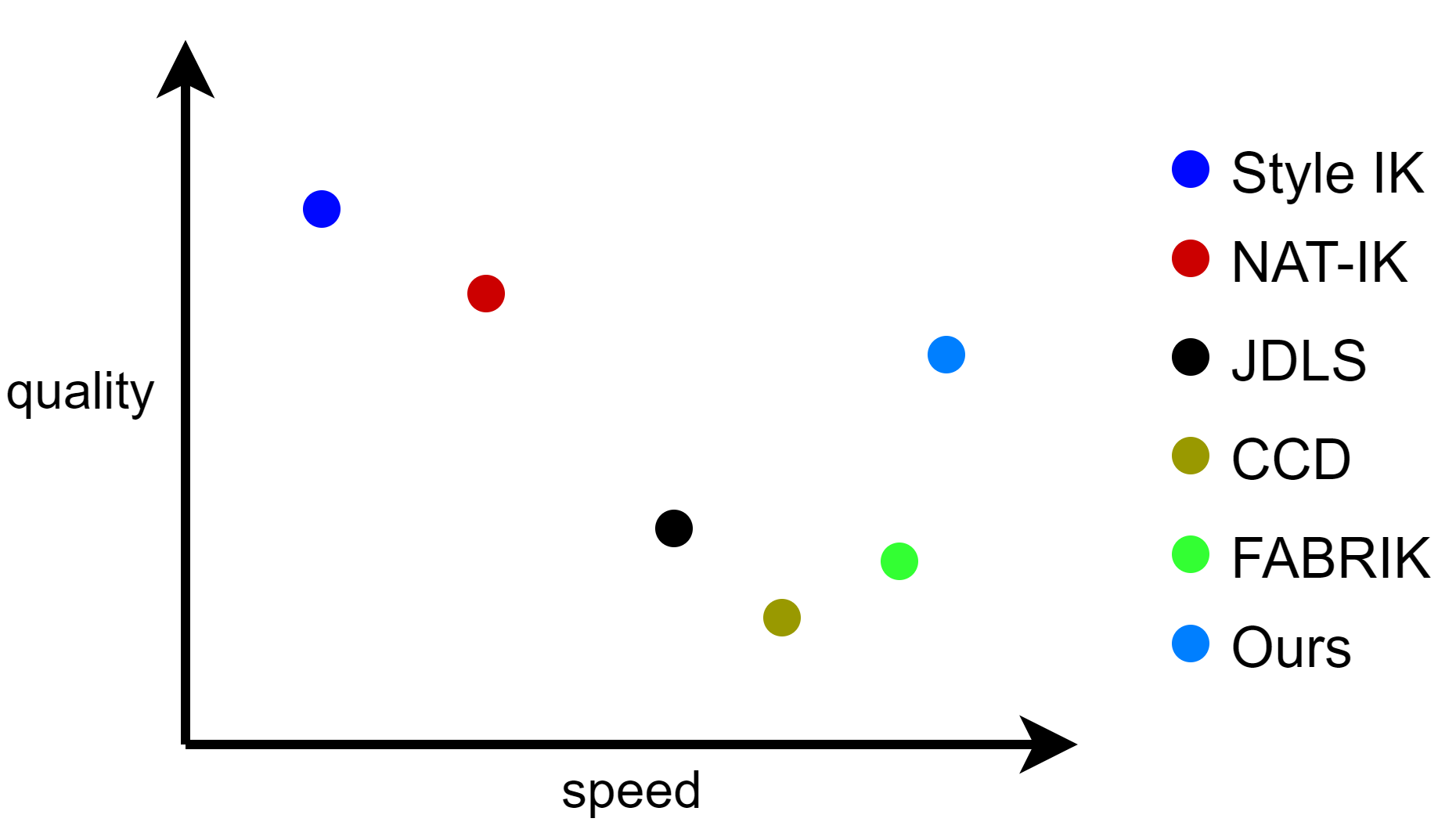}
    \caption{
    \begin{modified}
    General comparison of various full-body IK methods in terms of speed and quality.
    Style IK \cite{wu_posing_2009},
    NAT-IK \cite{Huang_IK_MGDM_2017},
    JDLS \cite{buss_selectively_2005},
    CCD \cite{wang_ccd_1991}
    FABRIK \cite{aristidou_fabrik:_2011}
    \end{modified}
    }
    \label{fig:comparisons}
\end{figure}
\end{modified}

%% file: conclusion.tex
We propose a method leveraging neural networks to provide an interactive and efficient tool to pose a character's skeleton. Learning from a large dataset of ground truth poses allows us to avoid manually specifying the complex constraints of the human skeleton, and only generating plausible poses.
Our approach also shifts a large part of the algorithmic burden of traditional methods to the training phase, allowing it to run competitively fast once set up. Compared to previous data-driven pose edition methods, our method takes up a small amount of memory, freeing up resources for other processes. We provide examples of integration of our method in a prototype posing software, as well as a way to switch to multiple targets configurations.
Future work on the subject will focus on extending the method to more use cases: using the method for other, non standardized skeletons (with a different morphology or non-humanoid) and adding joint rotations to the solver's input and output.